\documentstyle[12pt,epsf]{article}

\begin{document}

\begin{titlepage} 
\begin{center} 
 
{\Large \bf Spontaneous Symmetry Breaking in 
Compactified $\lambda\varphi^4$ Theory}\\
\vspace{.3in} 
{\large\em A.P.C. Malbouisson${}^{(a)}$, J.M.C. Malbouisson${}^{(b)}$,
A.E. Santana${}^{(b)}$} \\
\vspace{0.1cm} 
${}^{(a)}$ {\it Centro Brasileiro de Pesquisas F\'{\i}sicas,} \\ 
\vspace{0.1cm} 
{\it Rua Dr. Xavier Sigaud 150, 22290-180, Rio de Janeiro, RJ, Brazil } \\ 
\vspace{0.1cm} 
{\it E-mail: adolfo@cbpf.br} \\ 
\vspace{0.5cm} 
${}^{(b)}$ {\it Instituto de F\'{\i}sica, Universidade Federal da Bahia,} \\ 
\vspace{0.1cm} 
{\it Campus Universit\'ario de Ondina, 40210-340, Salvador, BA, Brazil} \\ 
\vspace{0.1cm}  
{\it E-mail: jmalboui@ufba.br, santana@fis.ufba.br} \\ 
\vspace{0.5cm}

\end{center} 
 
\subsection*{Abstract}   
We consider the massive  vector $N$-component $(\lambda\varphi^{4})_{D}$
theory in  Euclidian space and, using an extended Matsubara formalism we
perform a
compactification on a $d$-dimensional subspace, $d\leq D$. This 
allows us to treat jointly the effect of temperature and spatial confinement
in the effective potential of the model, setting forth grounds for an
analysis of  phase transitions driven by temperature and spatial boundary.
For $d=2$, which corresponds to the 
heated system confined between two parallel planes (separation $L$), we 
obtain, in the large $N$ limit at one-loop order, 
formulas for  temperature- and boundary-dependent  mass and coupling 
constant. The equation for the critical curve in the $\beta \times L$ plane
is also derived.

\end{titlepage} 
\newpage \baselineskip.37in

\section{Introduction}

Questions concerning stability and existence of phase transitions are
usually raised in the realm of field theories as an effect of temperature,
but similar queries may also appear for confined systems, if we consider the
behaviour of field theories as function of the spatial boundaries. One of
the most striking examples of this kind of situation is the Abrikosov
lattice of vortices occuring in type-II superconductors in high magnetic
fields \cite{Abrikosov}. A recent account on aspects of finite-temperature
field theory and on the behaviour of systems confined by spatial boundaries
can be found in Refs. \cite{NAdolfo1,NAdolfo2,Gino1,JMario}. In another
related domain of investigation, there are systems presenting defects
created, for instance, in the process of crystal growth by some prepared
circumstances (domain walls). At the level of effective field theories, in
many situations this can be modeled by considering a Dirac fermionic field
whose mass changes its sign as it crosses the defect. In other words, the
domain wall plays the role of a critical boundary, separating two different
states of the system (see for instance \cite{Cesar,FAdolfo} and the
references therein). In any case, the problem regarding to the behaviour of
physical parameters in field theories (masses or coupling constants) on
temperature as well as on spatial boundaries or inhomogeneities is still
largely open and deserves further investigation. These aspects are addressed
here by using a modified Matsubara scheme.

For Euclidian field theories, imaginary time (temperature) and the spatial
coordinates are on exactly the same footing, so that the Matsubara formalism
can be also applied for the breaking of invariance along any of the spatial
directions. Therefore, the existence of phase transitions induced either by
spatial boundaries (separating different states of the system) or by
temperature or both can be analyzed in a general fashion.

In order to advance with, in this note we present a general procedure to
compactify a $d$-dimensional subspace for the massive vector $N$-component $ 
(\lambda \varphi ^{4})_{D}$ theory in Euclidian space. This permits to
extending to an arbitrary subspace results in the literature for finite
temperature field theory \cite{NAdolfo1,NAdolfo2,Gino1} and for the
behaviour of field theories in presence of spatial boundaries \cite{JMario}.
Our aim is to construct a general framework to study effects associated to
simultaneous heating and spatial confinement, for a model that allows a
non-perturbative approach. We consider the finite-temperature vector $N$
-component $(\lambda \varphi ^{4})_{D}$ Euclidian theory at leading order in 
$\frac{1}{N}$, the system being submitted to the constraint of
compactification of a $d$-dimensional subspace, $d\leq D$. After describing
the general formalism, the renormalization is implemented for the
interesting particular case with $d=2$, $i.e.$, the heated system at a
temperature $\beta ^{-1}$ confined between two infinite parallel planes a
distance $L$ apart from one another. For this situation, we also derive the
equation for the critical curve in the $\beta \times L$ plane.

\section{Compactification of a $d$-dimensional subspace}

We consider the model described by the Lagrangian density, 
\begin{equation}
{\cal {L}}=\frac{1}{2}\partial _{\mu }\varphi _{a}\partial ^{\mu }\varphi
_{a}+\frac{1}{2}m^{2}\varphi _{a}\varphi _{a}+\frac{\lambda }{N}(\varphi
_{a}\varphi _{a})^{2},  \label{Lagrangeana}
\end{equation}
in Euclidian $D$-dimensional space, where $\lambda $ is the coupling
constant, $m_{0}$ is the mass (see below) and summation over repeated
``colour'' indices $a$ is assumed. To simplify the notation in the following
we drop out the colour indices, summation over them being understood in
field products. We consider the system in thermal equilibrium with a
reservoir at temperature $\beta ^{-1}$ and confined to a $(d-1)$-dimensional
spatial rectangular box of sides $L_{j}$, $j=2,3,...,d$. We use Cartesian
coordinates ${\bf r}=(x_{1},...,x_{d},{\bf z})$, where ${\bf z}$ is a $(D-d)$
-dimensional vector, with corresponding momentum ${\bf k}=(k_{1},...,k_{d}, 
{\bf q})$, ${\bf q}$ being a $(D-d)$-dimensional vector in momentum space.
Then we are allowed to introduce a generalized Matsubara prescription,
performing the following multiple replacements (compactification of a $d$
-dimensional subspace), 
\begin{equation}
\int \frac{dk_{i}}{2\pi }\rightarrow \frac{1}{L_{i}}\sum_{n_{i}=-\infty
}^{+\infty }\;;\;\;\;\;\;\;k_{i}\rightarrow \frac{2n_{i}\pi }{L_{i}} 
\;,\;\;i=1,2...,d,  \label{Matsubara1}
\end{equation}
where it is understood that one of the axis (say, $i=1$) corresponds to
imaginary time (inverse temperature). A simpler situation is the heated
system confined between two parallel planes a distance $L$ apart from one
another, normal to the $x$ -axis. In this case we have Cartesian coordinates 
${\bf r}=(\tau ,x,{\bf z})$, where formally $\tau $ correspond to the
imaginary time axis and ${\bf z}$ is a $(D-2)$-dimensional vector, with
corresponding momenta ${\bf k}=(k_{\tau },k_{x},{\bf q})$, ${\bf q}$ being a 
$(D-2)$-dimensional vector in momenta space. Then the multiple Matsubara
replacements above reduce to, 
\begin{equation}
\int \frac{dk_{\tau }}{2\pi }\rightarrow \frac{1}{\beta }\sum_{n_{1}=-\infty
}^{+\infty },\;\;\int \frac{dk_{x}}{2\pi }\rightarrow \frac{1}{L} 
\sum_{n_{2}=-\infty }^{+\infty }\;\;\;\;;\;\;\;k_{\tau }\rightarrow \frac{ 
2n_{1}\pi }{\beta }\;,\;\;k_{x}\rightarrow \frac{2n_{2}\pi }{L}.
\label{Matsubara2}
\end{equation}
Notice that compactification can be implemented in different ways, as for
instance by imposing specific conditions on the fields at spatial
boundaries. Here, we choose periodic boundary conditions in order to have a
unified Matsubara prescription for both temperature and spatial dimensions.

We start from the well known expression for the one-loop contribution to the
zero-temperature effective potential \cite{IZ}, 
\begin{equation}
U_{1}(\phi _{0})=\sum_{s=1}^{\infty }\frac{(-1)^{s+1}}{2s}\left[ \frac{ 
\lambda \phi _{0}^{2}}{2}\right] ^{s}\int \frac{d^{D}k}{(k^{2}+m^{2})^{s}},
\label{potefet0}
\end{equation}
where $m$ is the $renormalized$ mass. For the {\it {Wick-ordered}} model,
since in this case the tadpoles are suppressed, it is unnecessary to perform
an explicit mass renormalization at the order $\frac{1}{N}$ at the one-loop
approximation, the parameter $m_{0}$ in Eq.(\ref{Lagrangeana}) playing in
this case the role of the physical mass.

In the following, to deal with dimensionless quantities in the
regularization procedures, we introduce parameters $c^{2}=m^{2}/4\pi ^{2}\mu
^{2},\;\;(L_{i}\mu )^{2}=a_{i}^{-1},\;\;g=(\lambda /8\pi ^{2}),\;\;(\varphi
_{0}/\mu )=\phi _{0}$, where $\varphi _{0}$ is the normalized vacuum
expectation value of the field (the classical field) and $\mu $ is a mass
scale. In terms of these parameters and performing the Matsubara
replacements (\ref{Matsubara1}), the one-loop contribution to the effective
potential can be written in the form, 
\begin{equation}
U_{1}(\phi _{0},a_{1},...,a_{d})=\mu ^{D}\sqrt{a_{1}\cdots a_{d}}
\sum_{s=1}^{\infty }\frac{(-1)^{s}}{2s}g^{s}\phi
_{0}^{2s}\sum_{n_{1},...,n_{d}=-\infty }^{+\infty }\int \frac{d^{D-d}q}{
(a_{1}n_{1}^{2}+\cdots +a_{d}n_{d}^{2}+c^{2}+{\bf q}^{2})^{s}}.
\label{potefet1}
\end{equation}
Using a well-known dimensional regularization formula \cite{Zinn} to perform
the integration over the ($D-d$) non-compactfied momentum variables, we
obtain 
\begin{equation}
U_{1}(\phi _{0},a_{1},...,a_{d})=\mu ^{D}\sqrt{a_{1}\cdots a_{d}}
\sum_{s=1}^{\infty }f(D,d,s)g^{s}\phi _{0}^{2s}A_{d}^{c^{2}}(s-\frac{D-d}{2}
;a_{1},...,a_{d}),  \label{potefet2}
\end{equation}
where 
\begin{equation}
f(D,d,s)=\pi ^{(D-d)/2}\frac{(-1)^{s+1}}{2s\Gamma (s)}\Gamma (s-\frac{D-d}{2}
)
\end{equation}
and 
\begin{eqnarray}
A_{d}^{c^{2}}(\nu ;a_{1},...,a_{d}) &=&\sum_{n_{1},...,n_{d}=-\infty
}^{+\infty }(a_{1}n_{1}^{2}+\cdots +a_{d}n_{d}^{2}+c^{2})^{-\nu }=\frac{1}{
c^{2\nu }}  \nonumber  \label{zeta} \\
&&+2\sum_{i=1}^{d}\sum_{n_{i}=1}^{\infty }(a_{i}n_{i}^{2}+c^{2})^{-\nu
}+2^{2}\sum_{i<j=1}^{d}\sum_{n_{i},n_{j}=1}^{\infty
}(a_{i}n_{i}^{2}+a_{j}n_{j}^{2}+c^{2})^{-\nu }+\cdots  \nonumber \\
&&+2^{d}\sum_{n_{1},...,n_{d}=1}^{\infty }(a_{1}n_{1}^{2}+\cdots
+a_{d}n_{d}^{2}+c^{2})^{-\nu }.
\end{eqnarray}
Next we can proceed generalizing to several dimensions the mode-sum
regularization recipe described in Ref. \cite{Elizalde}. Using the identity, 
\begin{equation}
\frac{1}{\Delta ^{\nu }}=\frac{1}{\Gamma (\nu )}\int_{0}^{\infty }dt\;t^{\nu
-1}e^{-\Delta t},
\end{equation}
we get, 
\begin{eqnarray}
A_{d}^{c^{2}}(\nu ;a_{1},...,a_{d}) &=&\frac{1}{\Gamma (\nu )}
\int_{0}^{\infty }dt\;t^{\nu -1}e^{-c^{2}t}\left[ 1+2
\sum_{i=1}^{d}T_{1}(t,a_{i})+\right.  \nonumber  \label{zeta1} \\
&&\left. +2^{2}\sum_{i,j=1}^{d}T_{2}(t,a_{i},a_{j})+\cdots
+2^{d}T_{d}(t,a_{1},...,a_{d})\right] ,
\end{eqnarray}
where, 
\begin{eqnarray}
T_{1}(t,a_{i}) &=&\sum_{n_{i}=1}^{\infty }e^{-a_{i}n_{i}^{2}t}\;,  \label{T1}
\\
T_{j}(t,a_{1},...,a_{j})
&=&T_{j-1}(t,a_{1},...,a_{j-1})T_{1}(t,a_{j})\;\;\;\;,j=2,...,d.
\end{eqnarray}
Considering the property of functions $T_{1}$, 
\begin{equation}
T_{1}(t,a_{i})=-\frac{1}{2}+\sqrt{\frac{\pi }{a_{i}t}}\left[ \frac{1}{2}+S( 
\frac{\pi ^{2}}{a_{i}t})\right] ,  \label{T2}
\end{equation}
where 
\begin{equation}
S(x)=\sum_{n=1}^{\infty }e^{-n^{2}x},  \label{S}
\end{equation}
we can notice that the surviving terms in Eq.(\ref{zeta1}) are proportional
to $(a_{1}\cdots a_{d})^{-(1/2)}$. Therefore we find, 
\begin{eqnarray}
A_{d}^{c^{2}}(\nu ;a_{1},...,a_{d}) &=&\frac{\pi ^{\frac{d}{2}}}{\sqrt{
a_{1}\cdots a_{d}}}\frac{1}{\Gamma (\nu )}\int_{0}^{\infty }dt\;t^{(\nu - 
\frac{d}{2})-1}e^{-c^{2}t}  \nonumber  \label{zeta3} \\
&&\times \left[ 1+2\sum_{i=1}^{d}S(\frac{\pi ^{2}}{a_{i}t}
)+2^{2}\sum_{i<j=1}^{d}S(\frac{\pi ^{2}}{a_{i}t})S(\frac{\pi ^{2}}{a_{j}t}
)+\cdots +2^{d}\prod_{i=1}^{d}S(\frac{\pi ^{2}}{a_{i}t})\right] .
\end{eqnarray}
Inserting in Eq.(\ref{zeta3}) the explicit form of the function $S(x)$ in
Eq.(\ref{S}) and using the following representation for Bessel functions of
the third kind, $K_{\nu }$, 
\begin{equation}
2(a/b)^{\frac{\nu }{2}}K_{\nu }(2\sqrt{ab})=\int_{0}^{\infty }dx\;x^{\nu
-1}e^{-(a/x)-bx},  \label{K}
\end{equation}
we obtain after some long but straightforward manipulations, 
\begin{eqnarray}
A_{d}^{c^{2}}(\nu ;a_{1},...,a_{d}) &=&\frac{2^{\nu -\frac{d}{2}+1}\pi
^{2\nu -\frac{d}{2}}}{\sqrt{a_{1}\cdots a_{d}}\,\Gamma (\nu )}\left[ 2^{\nu
- \frac{d}{2}-1}\Gamma (\nu -\frac{d}{2})(\frac{m}{\mu })^{d-2\nu }\right. 
\nonumber  \label{zeta4} \\
&&\left. +2\sum_{i=1}^{d}\sum_{n_{i}=1}^{\infty }(\frac{m}{\mu
^{2}L_{i}n_{i} })^{\frac{d}{2}-\nu }K_{\nu -\frac{d}{2}}(mL_{i}n_{i})+\cdots
\right.  \nonumber \\
&&\left. +2^{d}\sum_{n_{1},...,n_{d}=1}^{\infty }(\frac{m}{\mu ^{2}\sqrt{
L_{1}^{2}n_{1}^{2}+\cdots +L_{d}^{2}n_{d}^{2}}})^{\frac{d}{2}-\nu }K_{\nu - 
\frac{d}{2}}(m\sqrt{L_{1}^{2}n_{1}^{2}+\cdots +L_{d}^{2}n_{d}^{2}})\right] .
\nonumber \\
&&
\end{eqnarray}
Taking $\nu =s-(D-d)/2$ in Eq.(\ref{zeta4}) we obtain from Eq.(\ref{potefet2}
) the one-loop correction to the effective potential in $D$ dimensions with
a compactified $d$-dimensional subspace, 
\begin{eqnarray}
U_{1}(\phi _{0},a_{1},...,a_{d}) &=&\mu ^{D}\sum_{s=1}^{\infty }g^{s}\phi
_{0}^{2s}h(D,s)\left[ 2^{s-\frac{D}{2}-2}\Gamma (s-\frac{D}{2})(\frac{m}{\mu 
})^{D-2s}\right.  \nonumber \\
&&\left. +\sum_{i=1}^{d}\sum_{n_{i}=1}^{\infty }(\frac{m}{\mu ^{2}L_{i}n_{i}}
)^{\frac{D}{2}-s}K_{\frac{D}{2}-s}(mL_{i}n_{i})\right.  \nonumber \\
&&\left. +2\sum_{i<j=1}^{d}\sum_{n_{i},n_{j}=1}^{\infty }(\frac{m}{\mu ^{2} 
\sqrt{L_{i}^{2}n_{i}^{2}+L_{j}^{2}n_{j}^{2}}})^{\frac{D}{2}-s}K_{\frac{D}{2}
-s}(m\sqrt{L_{i}^{2}n_{i}^{2}+L_{j}^{2}n_{j}^{2}})+\cdots \right.  \nonumber
\\
&&\left. +2^{d-1}\sum_{n_{1},\cdots n_{d}=1}^{\infty }(\frac{m}{\mu ^{2} 
\sqrt{L_{1}^{2}n_{1}^{2}+\cdots +L_{d}^{2}n_{d}^{2}}})^{\frac{D}{2}-s}K_{ 
\frac{D}{2}-s}(m\sqrt{L_{1}^{2}n_{1}^{2}+\cdots +L_{d}^{2}n_{d}^{2}})\right]
,  \nonumber \\
&&  \label{potefet3}
\end{eqnarray}
with 
\begin{equation}
h(D,s)=\frac{1}{2^{D/2-s-1}\pi ^{D/2-2s}}\frac{(-1)^{s+1}}{s\Gamma (s)}.
\label{h}
\end{equation}
For $d=2$, with $L_{1}=\beta $ and $L_{2}=L$, we get the thermal and
boundary dependent one-loop correction to the effective potential, 
\begin{eqnarray}
U_{1}(\phi _{0},\beta ,L) &=&\mu ^{D}\sum_{s=1}^{\infty }g^{s}\phi
_{0}^{2s}h(D,s)\left[ 2^{s-\frac{D}{2}-2}\Gamma (s-\frac{D}{2})(\frac{m}{\mu 
})^{D-2s}\right.  \nonumber \\
&&\left. +\sum_{n=1}^{\infty }(\frac{m}{\mu ^{2}\beta n})^{\frac{D}{2}-s}K_{ 
\frac{D}{2}-s}(m\beta n)+\sum_{n=1}^{\infty }(\frac{m}{\mu ^{2}Ln})^{\frac{D 
}{2}-s}K_{\frac{D}{2}-s}(mLn)\right.  \nonumber \\
&&\left. +2\sum_{n_{1},n_{2}=1}^{\infty }(\frac{m}{\mu ^{2}\sqrt{\beta
^{2}n_{1}^{2}+L^{2}n_{2}^{2}}})^{\frac{D}{2}-s}K_{\frac{D}{2}-s}(m\sqrt{
\beta ^{2}n_{1}^{2}+L^{2}n_{2}^{2}})\right] .  \label{potefet4}
\end{eqnarray}
For $d=1$, Eq.(\ref{potefet3}) reduces to the cases of thermal or boundary
dependences separately, studied in Refs. \cite{NAdolfo1}, \cite{Gino1} and 
\cite{JMario}. In the next section we apply the formalism presented here to
the simpler case $d=2$: the heated system at a temperature $\beta ^{-1}$
confined between two parallel planes a distance $L$ apart from one another.

\section{Renormalization scheme}

We consider in the following the zero external-momenta four-point function,
which is the basic object for our definition of the renormalized coupling
constant. The four-point function at leading order in $\frac{1}{N}$ is given
by the sum of all diagrams of the type depicted in $Fig.1$. This sum gives
for the $\beta$ and $L$ -dependent four-point function at zero external
momenta the formal expression, 
\begin{equation}
\Gamma _{D}^{(4)}(0,\beta ,L)=\frac{1}{N}\;\;\frac{\lambda }{1-\lambda
\Sigma (D,\beta,L)},  \label{4-point1}
\end{equation}
where $\Sigma (D,\beta,L)$ corresponds to the single bubble subdiagram in $ 
Fig.1$. To obtain an expression for $\Sigma (D,\beta,L)$, we generalize
one-loop results from finite temperature and confined field theory (\cite
{NAdolfo2}, \cite{Gino1}, \cite{JMario}) to our present situation. These
results are obtained by the concurrent use of dimensional and $zeta$
-function analytic regularizations, to evaluate formally the integral over
the continuous momenta and the summation over the Matsubara frequencies. We
get sums of polar terms plus analytic corrections depending on the
compactified variables. Renormalized quantities are obtained by subtraction
of the divergent (polar) terms appearing in the quantities obtained by
application of the modified Feynman rules (Matsubara prescriptions) and
dimensional regularization formulas. These polar terms are proportional to $ 
\Gamma$-functions having the dimension $D$ in the argument and correspond to
the introduction of counter terms in the original Lagrangian density. In
order to have a coherent procedure in any dimension, these subtractions
should be performed even in the case of odd dimension $D$, where no poles of 
$\Gamma $-functions are present.

Note that since we are using dimensional regularization techniques, there is
implicitly in the above formulas a factor $\mu ^{4-D}$ in the definition of
the coupling constant. In what follows we make explicit this factor, the
symbol $\lambda $ standing for the dimensionless coupling parameter (which
coincides with the physical coupling constant in $D=4$). To proceed we use
the renormalization conditions, 
\begin{equation}
\frac{\partial ^{2}}{\partial \phi ^{2}}U_{1}(D,\beta ,L)|_{\phi
_{0}=0}=m^{2}\mu ^{2}  \label{renorm1}
\end{equation}
and 
\begin{equation}
\frac{\partial ^{4}}{\partial \phi ^{4}}U_{1}(D,\beta ,L)_{\phi
_{0}=0}=\lambda \mu ^{4},  \label{renorm2}
\end{equation}
from which we deduce that formally the single bubble function $\Sigma
(D,\beta ,L)$ is obtained from the coefficient of the fourth power of the
field ($s=2$) in Eq. (\ref{potefet4}). Of course such coefficient is in
general ultraviolet divergent and a renormalization procedure is needed.
Then using Eqs. (\ref{renorm2}) and (\ref{potefet4}) we can write $\Sigma
(D,\beta ,L)$ in the form, 
\begin{equation}
\Sigma (D,\beta ,L)=H(D)-G(D,\beta ,L),  \label{Sigma}
\end{equation}
where the $\beta $ and $L$ dependent contribution $G(D,\beta ,L)$ comes from
the second term between brackets in Eq.(\ref{potefet4}), that is, 
\begin{eqnarray}
G(D,\beta ,L) &=&\frac{3}{2}\frac{\mu ^{4-D}}{(2\pi )^{D/2}}\left[
\sum_{n=1}^{\infty }(\frac{m}{n\beta })^{\frac{D}{2}-2}K_{\frac{D}{2}
-2}(n\beta m)+\sum_{n=1}^{\infty }(\frac{m}{nL})^{\frac{D}{2}-2}K_{\frac{D}{
2 }-2}(nLm)\right.  \nonumber  \label{G} \\
&&\left. +2\sum_{n_{1},n_{2}=1}^{\infty }(\frac{m}{\sqrt{\beta
^{2}n_{1}^{2}+L^{2}n_{2}^{2}}})^{\frac{D}{2}-2}K_{\frac{D}{2}-2}(m\sqrt{
\beta ^{2}n_{1}^{2}+L^{2}n_{2}^{2}})\right],
\end{eqnarray}
and $H(D)$, a polar parcel coming from the first term between brackets in
Eq.(\ref{potefet4}), is 
\begin{equation}
H(D)\propto \Gamma (2-\frac{D}{2})(m/\mu )^{D-4}.  \label{H}
\end{equation}
We see from Eq.(\ref{H}) that for even dimensions $D\geq 4$, $H(D)$ is
divergent, due to the pole of the $\Gamma $-function. Accordingly this term
must be subtracted to give the renormalized single bubble function $\Sigma
_{R}(D,\beta ,L)$. We get simply, 
\begin{equation}
\Sigma _{R}(D,\beta ,L)=-G(D,\beta ,L).  \label{sigmaR}
\end{equation}
As mentioned before, for sake of uniformity the term $H(D)$ is also
subtracted in the case of odd dimension $D$, where no poles of $\Gamma $
-functions are present. In these cases we perform a finite renormalization.
>From the properties of Bessel functions it can be seen from Eq.(\ref{G})
that for any dimension $D$, $G(D,\beta ,L)$ satisfies the conditions, 
\begin{equation}
\lim_{\beta ,L\rightarrow \infty }G(D,\beta
,L)=0\;,\;\;\;\;\;\;\;\lim_{\beta ,L\rightarrow 0}G(D,\beta ,L)\rightarrow
\infty ,  \label{GG1}
\end{equation}
and $G(D,\beta ,L) > 0$ for any values of $D$, $\beta $ and $L$.

Let us define the $\beta$ and $L$ dependent renormalized coupling constant $ 
\lambda_{R}(D,\beta,L)$ at the leading order in $1/N$ as, 
\begin{equation}
\Gamma _{D}^{(4)}(0,\beta,L)\equiv \frac{1}{N}\lambda_{R}(D,\beta,L)=\frac{1 
}{N}\;\;\frac{\lambda}{1-\lambda \Sigma_{R}(D,\beta,L)}  \label{lambR}
\end{equation}
and the renormalized zero-temperature coupling constant in the absence of
boundaries as, 
\begin{equation}
\frac{1}{N}\lambda_{R}(D) =\lim_{\beta,L\rightarrow \infty }\Gamma
_{D,R}^{(4)}(0,\beta,L).  \label{lambfree}
\end{equation}
>From Eqs.(\ref{lambfree}), (\ref{lambR}), (\ref{GG1}) and (\ref{sigmaR}) we
conclude that $\lambda_{R}(D)=\lambda$. In other words we have done a choice
of renormalization scheme such that the constant $\lambda$ introduced in the
Lagrangian corresponds to the physical coupling constant in free space, at
zero temperature. From Eqs.(\ref{lambR}) and (\ref{sigmaR}) we obtain the $ 
\beta$ and $L$ dependent renormalized coupling constant, 
\begin{equation}
\lambda_{R}(D,\beta,L)=\frac{\lambda}{1+\lambda G(D,\beta,L)}.
\label{lambdaR1}
\end{equation}
The above procedure corresponds, on perturbative grounds, to sum up all the
chains of bubble graphs in $Fig.1$. It is just the resummation of all the
perturbative contributions including the counter terms from the chains of
bubbles and the subtraction of the divergent (polar) parts, written in
compact form. Again, these subtractions should be performed even for odd
dimensions.

We have used above a modified minimal subtraction scheme where the mass and
coupling constant counter terms are poles at the physical values of $s$. The
multiple and double sums in Eqs. (\ref{zeta}) to (\ref{potefet4}) define
generalized inhomogeneous Epstein-Hurwitz $zeta$-functions, in such a way
that the $\beta $ and $L$ dependent correction to the coupling constant is
proportional to the regular part of the analytical extension of these
inhomogeneous Epstein-Hurwitz $zeta$-functions in the neighbourhood of the
pole at $s=2$. For the {\it {non-Wick-ordered}} model the same argument
applies to the renormalized mass for $s=1$, and the $\beta $ and $L$
dependent {\it renormalized} mass at one-loop approximation is given by, 
\begin{eqnarray}
m^{2}(\beta ,L) &=&m_{0}^{2}+\frac{4\mu ^{4-D}\lambda }{(2\pi )^{D/2}}\left[
\sum_{n=1}^{\infty }(\frac{m}{n\beta })^{\frac{D}{2}-1}K_{\frac{D}{2}
-1}(n\beta m)+\sum_{n=1}^{\infty }(\frac{m}{nL})^{\frac{D}{2}-1}K_{\frac{D}{
2 }-1}(nLm)\right.  \nonumber  \label{mDyson} \\
&&\left. +2\sum_{n_{1},n_{2}=1}^{\infty }(\frac{m}{\sqrt{\beta
^{2}n_{1}^{2}+L^{2}n_{2}^{2}}})^{\frac{D}{2}-1}K_{\frac{D}{2}-1}(m\sqrt{
\beta ^{2}n_{1}^{2}+L^{2}n_{2}^{2}})\right] .
\end{eqnarray}

In dimension $D=3$, from the relation \cite{Abramowitz}, 
\begin{equation}
K_{\frac{1}{2}}(z)= \sqrt{\frac{\pi }{2z}}e^{-z},  \label{Abramov}
\end{equation}
after summing up geometric series, we obtain the following semi closed
expressions, 
\begin{equation}
G(\beta,L)=\frac{3\mu}{8\pi m}\left[(e^{m\beta}-1)^{-1}+(e^{mL}-1)^{-1}+2
\sum_{n_{1},n_{2}=1}^{\infty}e^{-m\sqrt{\beta^{2}n_{1}^{2}+L^{2}n_{2}^{2}}}  
\right] ,  \label{G3}
\end{equation}

\begin{equation}
m^{2}(\beta,L)=m_{0}^{2}+\frac{\mu \lambda}{\pi}\left[-\frac{
\log(1-e^{-m\beta})}{\beta}-\frac{\log(1-e^{-mL})}{L}+2
\sum_{n_{1},n_{2}=1}^{\infty}\frac{e^{-m\sqrt{
\beta^{2}n_{1}^{2}+L^{2}n_{2}^{2}}}}{\sqrt{\beta^{2}n_{1}^{2}+L^{2}n_{2}^{2}}
}\right].  \label{m3}
\end{equation}

\section{Mass behaviour and critical curve}

As far as the mass behaviour is concerned, if we start in the ordered phase
with a negative squared mass, the model exhibits spontaneous symmetry
breaking of the $O(N)$ symmetry to $O(N-1)$, but for sufficiently small
critical values of $\beta $ and $L$ the symmetry is restored. We can define
the critical curve $C(\beta _{c},L_{c})$ as the curve in the $\beta \times L$
plane for which the inverse squared correlation length , $\xi ^{-2}(\beta
,L,\varphi _{0})$, vanishes in the gap equation, 
\begin{eqnarray}
\xi ^{-2}(\beta ,L,{\bf \varphi }_{0}) &=&m_{0}^{2}+\lambda \varphi _{0}^{2}
\nonumber  \label{gap} \\
&&+\frac{\lambda (N+2)}{2N\beta L}\sum_{n_{1},n_{2}=-\infty }^{\infty }\int 
\frac{d^{D-2}q}{(2\pi )^{D-2}}\;\frac{1}{{\bf q}^{2}+(\frac{2\pi n_{1}}{
\beta })^{2}+(\frac{2\pi n_{2}}{L})^{2}+\xi ^{-2}(\beta ,L,\varphi _{0})}, 
\nonumber \\
&&
\end{eqnarray}
where ${\bf \varphi }_{0}$ is the normalized vacuum expectation value of the
field (different from zero in the ordered phase).

Close to the critical curve, $\varphi _{0}$ vanishes and the thermal gap
equation at one-loop order reduces, in the large $N$ limit, to Eq.(\ref
{mDyson}). If we limit ourselves to the neighbourhood of criticality, $ 
m^{2}\approx 0$, we may use an asymptotic formula for small values of the
argument of Bessel functions, 
\begin{equation}
K_{\nu }(z)\approx \frac{1}{2}\Gamma (\nu )\left( \frac{z}{2}\right) ^{-\nu
}\;\;\;(z\sim 0)  \label{K}
\end{equation}
and Eq.(\ref{mDyson}) becomes, 
\begin{equation}
m^{2}(\beta ,L)\approx m_{0}^{2}+\frac{4\lambda \mu ^{4-D}}{(\pi )^{D/2}} 
\Gamma (\frac{D}{2}-1)\left[ (\beta ^{2-D}+L^{2-D})\zeta
(D-2)+2E_{2}((D-2)/2;\beta ^{2},L^{2})\right] \;,  \label{mDysoncr}
\end{equation}
where $\zeta (D-2)$ is the Riemann $zeta$-function, 
\begin{equation}
\zeta (D-2)=\sum_{n=1}^{\infty }\frac{1}{n^{D-2}}
\end{equation}
and $E_{2}((D-2)/2;\beta ^{2},L^{2})$ is the generalized Epstein-Hurwitz $ 
zeta$ -function \cite{Kirsten}, 
\begin{equation}
E_{2}((D-2)/2;\beta ^{2},L^{2})=\sum_{n_{1},\,n_{2}=1}^{\infty }\frac{1}{ 
(\beta ^{2}n_{1}^{2}+L^{2}n_{2}^{2})^{\frac{D-2}{2}}}\,.  \label{Z}
\end{equation}
The Riemann $zeta$-function $\zeta (D-2)$ has an analytical extention to the
whole complex $D$-plane, having an unique simple pole (of residue $1$) at $ 
D=3$. On the other hand, from the properties of generalized Epstein $zeta$
-functions, in particular recurrence relations, we can write the generalized 
$zeta$-function (\ref{Z}) in terms of the Riemann $zeta$-function. As it is
done in Ref. \cite{Kirsten}, whatever the sum one chooses to perform
firstly, the manifest $\beta \leftrightarrow L$ symmetry of Eq. (\ref{Z}) is
lost. In order to preserve this symmetry, we adopt here a symmetrized
summation which leads to 
\begin{eqnarray}
E_{2}((D-2)/2;\beta ^{2},L^{2}) &=&-\frac{1}{4}\left( \beta
^{2-D}+L^{2-D}\right) \zeta (D-2)  \nonumber \\
&&+\frac{\sqrt{\pi }\Gamma (\frac{D-3}{2})}{4\Gamma (\frac{D-2}{2})}\left( 
\frac{1}{\beta L^{D-3}}+\frac{1}{\beta ^{D-3}L}\right) \zeta (D-3)  \nonumber
\\
&&+\frac{\sqrt{\pi }}{\Gamma (\frac{D-2}{2})}W_{2}(D-2;\beta ,L)\;,
\label{Z1}
\end{eqnarray}
where 
\begin{eqnarray}
W_{2}(D-2;\beta ,L) &=&\frac{1}{\beta ^{D-2}}\sum_{n_{1},n_{2}=1}^{\infty
}(\pi \frac{L}{\beta }n_{1}n_{2})^{\frac{D-3}{2}}K_{\frac{D-3}{2}}(2\pi 
\frac{L}{\beta }n_{1}n_{2})  \nonumber \\
&&+\frac{1}{L^{D-2}}\sum_{n_{1},n_{2}=1}^{\infty }(\pi \frac{\beta }{L} 
n_{1}n_{2})^{\frac{D-3}{2}}K_{\frac{D-3}{2}}(2\pi \frac{\beta }{L} 
n_{1}n_{2})\;,  \label{W}
\end{eqnarray}
from which we conclude that $E_{2}((D-2)/2;\beta ^{2},L^{2})$ has only two
simple poles at $D=4$ and $D=3$. The solution of Eq.(\ref{mDysoncr}) for $ 
m(\beta ,L)=0$ and $m_{0}^{2}<0$ defines the critical curve $C(\beta
_{c},L_{c})$ in Euclidean space dimension $D$ ($D\neq 4$ and $D\neq 3$), 
\begin{equation}
m_{0}^{2}+\frac{4\lambda \mu ^{4-D}}{(\pi )^{D/2}}\Gamma (\frac{D}{2}-1) 
\left[ (\beta _{c}^{2-D}+L_{c}^{2-D})\zeta (D-2)+2E_{2}((D-2)/2;\beta
_{c},L_{c})\right] =0.  \label{mDysoncr1}
\end{equation}

For $D=4$ the generalized $zeta$-function $Z_{2}$ has a pole and for $D=3$
both the Riemann $zeta$-function and the function $Z_{2}$ have poles. We can
not obtain a critical curve in dimensions $D=4$ and $D=3$ by a limiting
procedure from Eq.(\ref{mDysoncr1}). For $D=4$, which corresponds to the
physical interesting situation of the heated system confined between two
palallel planes embedded in a $3$-dimensional Euclidean space, Eqs.(\ref
{mDysoncr}) and (\ref{mDysoncr1}) become meaningless. To obtain a critical
curve in $D=4$, a regularization procedure can be done as follows: we
perform an analytic continuation of the $zeta$-function to values of the
argument $z\leq 1$, by means of the formula, 
\begin{equation}
\zeta (z)=\frac{1}{\Gamma (z/2)}\Gamma (\frac{1-z}{2})\pi ^{z-\frac{1}{2} 
}\zeta (1-z)\;.  \label{extensao}
\end{equation}
Eq.(\ref{extensao}) defines a meromorphic function having only one simple
pole at $z=1$. Next, remembering the formula, 
\begin{equation}
\lim_{z\rightarrow 1}\left[ \zeta (z)-\frac{1}{1-z}\right] ={\bf C\;},
\label{extensao1}
\end{equation}
where ${\bf C}$ is the Euler-Mascheroni constant, and using Eq. (\ref{Z1}),
we define the {\it renormalized} mass 
\begin{equation}
\lim_{D\rightarrow 4\_}\left[ m_{0}^{2}+\frac{1}{D-4}\frac{4\lambda }{\pi
^{2}\beta L}\right] =m^{2}\;,  \label{massren}
\end{equation}
in terms of which we obtain the critical curve in dimension $D=4$, 
\begin{equation}
m^{2}+\frac{\lambda }{3}\left( \frac{1}{\beta _{c}^{2}}+\frac{1}{L_{c}^{2}} 
\right) +\frac{\pi {\bf C}}{\beta _{c}L_{c}}+4\sqrt{\pi }W_{2}(2;\beta
_{c},L_{c})=0\;,  \label{mDysoncr2}
\end{equation}
where we have used $\zeta (2)=\pi ^{2}/6$ and $\Gamma (1/2)=\sqrt{\pi }$.

\section{Concluding remarks}

In this paper we have presented a generalization of the Matsubara formalism
to account for simultaneous compactification of imaginary-time and spatial
coordinates. The method is applied to the massive vector N-component $ 
(\lambda \varphi ^{4})_{D}$ theory, in Euclidian space, compactified in a $d$
-dimensional subspace ($d\leq D$), including temperature. We have calculated
the effective potential of the model from which, in the particular case $d=2$
, equations for the $\beta $- and $L$-dependent mass and coupling constant
have been derived. In the absence of Wick-ordering, the renormalized mass
can not be taken as the coefficient of the term $\varphi _{a}\varphi _{a}$
in the Lagrangian, and we must take a $\beta $- and $L$- corrected
renormalized mass, $m(\beta ,L)$. To obtain the $\beta $- and $L$- dependent
coupling constant, the mass parameter $m$ should be replaced by $m(\beta ,L)$
in Eqs( \ref{lambdaR1}) and (\ref{G}) and the resulting system of equations
should be solved. Under these conditions exact closed expressions are almost
impossible to obtain, since it would require a procedure equivalent to
solving exactly the Dyson-Schwinger equations. For $D=4$, an equation for
the critical curve in the $\beta \times L$ plane is obtained. An application
of the formalism presented here to perform a detailed analytical and
numerical study of the behaviour of the coupling constant and the mass on
temperature and on spatial boundaries is in progress and will be presented
elsewhere.

\section{Acknowledgments}

This paper was supported by CNPq (Brazilian National Research Council) and
FAPERJ (Foundation for the support of research in the state of Rio de
Janeiro - Brazil). One of us (A.P.C.M.) is grateful for kind hospitality to
Instituto de F\'{\i}isica da UFBA (Brazil) where part of this work has been
done. J.M.C.M. and A.E.S. thank CBPF/MCT for hospitality.

\bigskip \newpage

{\bf Figure captions}

Fig.1 - Typical diagram contributing to the four-point function at leading
order in $\frac{1}{N}$. To each vertex there is a factor $\frac{\lambda }{N}$
and for each single bubble a colour circulation factor $N$.


\begin{thebibliography}{99}
\bibitem{Abrikosov}  A.A. Abrikosov, Zh. Eskp. Teor. Fiz. {\bf 32}, 1442
(1957).

\bibitem{NAdolfo1}  A.P.C. Malbouisson, N.F. Svaiter, J. Math. Phys. {\bf 37}
, 4352 (1996).

\bibitem{NAdolfo2}  A.P.C. Malbouisson, N.F. Svaiter, Physica A {\bf 233},
573 (1996).

\bibitem{Gino1}  G.N.J. A\~{n}a\~{n}os, A.P.C. Malbouisson, N.F. Svaiter,
Nucl. Phys. B {\bf 547}, 221 (1999).

\bibitem{JMario}  A.P.C. Malbouisson, J.M.C. Malbouisson, J. Phys. A: Math.
Gen. {\bf 35}, 2263 (2002).

\bibitem{IZ}  C. Itzykson, J-B. Zuber, {\it Quantum Field Theory}
(McGraw-Hill, New York, 1980), p. 451.

\bibitem{Cesar}  C.D. Fosco, A. Lopez, Nucl. Phys. B {\bf 538}, 685 (1999).

\bibitem{FAdolfo}  L. Da Rold, C.D. Fosco, A.P.C. Malbouisson, Nucl. Phys. B 
{\bf 624}, 485 (2002).

\bibitem{Zinn}  J. Zinn-Justin, {\it Quantum Field Theory and Critical
Phenomena} (Clarendon Press, Oxford, 1996).

\bibitem{Elizalde}  A. Elizalde, E. Romeo, J. Math. Phys. {\bf 30}, 1133
(1989).

\bibitem{Abramowitz}  M. Abramowitz, I.A. Stegun, eds., {\it Handbook of
Mathematical Functions} (Dover, New York, 1965).

\bibitem{Kirsten}  K. Kirsten, J. Math. Phys. {\bf 35}, 459 (1994).
\end{thebibliography}
\end{document}